\documentclass[floats,floatfix,showpacs,amssymb,prd,superscriptaddress,twocolumn,aps]{revtex4}
\usepackage{graphicx, epsfig, amssymb} 
\usepackage{amsmath, amsfonts}
 \usepackage{gensymb}
\usepackage{bm} 

\usepackage[linktocpage]{hyperref}
\usepackage[caption=false]{subfig}
\usepackage[usenames]{color}

\def\be{\begin{equation}}
\def\ee{\end{equation}}
\def\beq{\begin{eqnarray}}
\def\eeq{\end{eqnarray}}

\newcommand{\bea}{\begin{eqnarray}}
\newcommand{\eea}{\end{eqnarray}}
\newcommand{\ben}{\begin{enumerate}}
\newcommand{\een}{\end{enumerate}}
\newcommand{\bi}{\begin{itemize}}
\newcommand{\ei}{\end{itemize}}

\pdfoutput=1

\begin{document}

\title{Spontaneous scalarisation of charged black holes}

 \author{Carlos~A.~R.~Herdeiro}
   \affiliation{
   Departamento de F\'\i sica da Universidade de Aveiro and CIDMA, 
   Campus de Santiago, 3810-183 Aveiro, Portugal.
 }

 \author{Eugen Radu}
   \affiliation{
   Departamento de F\'\i sica da Universidade de Aveiro and CIDMA, 
   Campus de Santiago, 3810-183 Aveiro, Portugal.
 }
 
 \author{Nicolas Sanchis-Gual}
\affiliation{Departamento de
  Astronom\'{\i}a y Astrof\'{\i}sica, Universitat de Val\`encia,
  Dr. Moliner 50, 46100, Burjassot (Val\`encia), Spain}

\author{Jos\'e A. Font}
\affiliation{Departamento de
  Astronom\'{\i}a y Astrof\'{\i}sica, Universitat de Val\`encia,
  Dr. Moliner 50, 46100, Burjassot (Val\`encia), Spain}
\affiliation{Observatori Astron\`omic, Universitat de Val\`encia, C/ Catedr\'atico 
  Jos\'e Beltr\'an 2, 46980, Paterna (Val\`encia), Spain}


\date{June 2018}

\begin{abstract}
Extended scalar-tensor-Gauss-Bonnet (eSTGB) gravity has been recently argued to exhibit spontaneous scalarisation of vacuum black holes (BHs). A similar phenomenon can be expected in a larger class of models, which includes $e.g$ Einstein-Maxwell-scalar (EMS) models, where spontaneous scalarisation of electrovacuum  BHs should occur. EMS models have no higher curvature corrections, a technical simplification over eSTGB models that allows us to investigate,  \textit{fully non-linearly}, BH scalarisation in two novel directions. Firstly, numerical simulations in spherical symmetry show, dynamically, that  Reissner-Nordstr\"om  (RN) BHs evolve into a perturbatively stable scalarised BH. Secondly, we compute the non-spherical sector of static scalarised BH solutions bifurcating from the RN BH trunk. Scalarised BHs form an infinite (countable) number of branches, and possess a large freedom in their multipole structure. Unlike the case of electrovacuum, the EMS model admits  static, asymptotically flat, regular on and outside the horizon BHs without spherical symmetry and even without any spatial isometries, which are thermodynamically preferred over the electrovacuum state. We speculate on a possible dynamical role of these non-spherical scalarised BHs. 
\end{abstract}




\maketitle
\noindent{\bf {\em Introduction.}} 
The newborn field of gravitational wave astronomy~\cite{Abbott:2016blz,Abbott:2016nmj,Abbott:2017vtc,Abbott:2017oio,TheLIGOScientific:2017qsa,Abbott:2017gyy} will test the nature of astrophysical black holes (BHs) in an unprecedented way. It is therefore of the utmost importance to theoretically scrutinise all physically reasonable alternatives to the Kerr BH~\cite{Kerr:1963ud} hypothesis, a paradigm  supported by the uniqueness theorems for vacuum General Relativity (GR)~\cite{Carter:1971zc,Robinson:1975bv,Chrusciel:2012jk}.

A reasonable alternative BH model requires, conservatively, being a solution of a consistent physical theory and having robust dynamical properties. The latter include  a formation mechanism and sufficient stability. A novel class of such models has recently emerged in extended scalar-tensor-Gauss-Bonnet (eSTGB) gravity~\cite{Doneva:2017bvd,Silva:2017uqg,Antoniou:2017acq}. In eSTGB the Schwarzschild BH may become spontaneously scalarised, since linear theory reveals it becomes unstable against scalar perturbations for sufficiently small BHs. 
Moreover,  BHs with scalar hair exist in the model which are thermodynamically preferred over the vacuum solution~\cite{Doneva:2017bvd} and stable against spherical perturbations~\cite{Blazquez-Salcedo:2018jnn}. This phenomenon is akin to the spontaneous scalarisation found long ago for neutron stars in the context of scalar-tensor theories~\cite{Damour:1993hw}, but with the scalarisation induced by strong spacetime curvature, rather than matter (see~\cite{Cardoso:2013fwa} for a discussion of matter induced BH scalarisation).

The studies of BH scalarisation in~\cite{Doneva:2017bvd,Silva:2017uqg,Antoniou:2017acq,Blazquez-Salcedo:2018jnn} were both non-dynamical and, for the non-linear scalarised solutions, restricted to the spherically symmetric sector. However, the instability of the vacuum solutions observed in linear theory contains non-spherical modes. This raises the issues of what other scalarised, static, BH solutions are admitted in eSTGB and what is the dynamical evolution and endpoint of the instability of vacuum BHs. 

A complete understanding of these evolutions requires non-linear  numerical simulations; these are challenging, particularly if no symmetries are imposed, in view of the higher curvature terms in eSTGB gravity. In this letter we tackle these questions by observing that the eSTGB model belongs to a larger universality class, that in particular includes the technically simpler Einstein-Maxwell-scalar (EMS) models~\footnote{See, $e.g.$,~\cite{Gubser:2005ih,Stefanov:2007eq,Doneva:2010ke} for earlier discussions of charged BHs scalarisation in different models.}. In an illustrative EMS model, we perform, in spherical symmetry, non-linear numerical simulations exhibiting, dynamically, the scalarisation of the Reissner-Nordstr\"om (RN) BH. The process indeed forms a perturbatively stable scalarised BH.  In the same EMS model, we study the non-spherical, static, scalarised solutions, and show they are thermodynamically preferred over the (electrovacuum) spherical BH.  Since in cousin models to EMS,  non-linear numerical evolutions of binary BHs have been performed~\cite{Hirschmann:2017psw}, even non-spherical, non-linear evolutions of the scalarisation instability in EMS models are within reach.

Besides exhibiting, dynamically, the scalarisation instabilty in the spherical sector, our investigation of the static non-spherical sector constructs the first examples of static, asymptotically flat, regular on and outside the event horizon BHs without spatial isometries. This is a maximal violation of Israel's uniqueness theorem~\cite{Israel:1967za}, stating that static BHs in electrovacuum must be spherical. The non-minimal couplings in EMS models circumvent in a radical way the uniqueness and simplicity of  (electro)vacuum BHs~\cite{Chrusciel:2012jk}, which holds even if a minimal coupling to a real scalar field is included~\cite{Bekenstein:1995un,Herdeiro:2015waa}. It endows BHs with~\textit{multipole freedom}.

\noindent{\bf {\em Universal conditions and a toy model.}} 
Consider the following scalar field action on a non-trivial background ($e.g.$ a curved spacetime and/or electromagnetic field):
\begin{eqnarray}
\label{actionS}
\mathcal{S}_\phi= -\int d^4 x \sqrt{-g} 
\left[
 2\partial_\mu \phi\partial^\mu \phi 
+f(\phi) {\cal I}(\psi;g)
\right] \ ;
\end{eqnarray}
$f(\phi) $ is the 
{\it coupling function}
and 
${\cal I}$ is a source term which generically depends on matter field(s),
$\psi$,
and the metric tensor, $g_{\mu\nu}$.
The scalar field equation of motion is $\Box \phi= f' {\cal I}/4$.  $f(\phi)$ must allow the existence of a \textit{scalar-free} solution, with $\phi=0$. This requires $f'(0)=0$.  Spontaneous scalarisation occurs if the scalar-free solution is unstable against scalar perturbations $\delta \phi$. These obey  $(\Box-\mu_{\rm eff}^2)\delta \phi =0$, where $\mu_{\rm eff}^2\equiv    f''(0) {\cal I}({\rm background})/4$. If $\mu_{\rm eff}^2<0$, 
a tachyonic instability sets in, driving the system away from the scalar-free solution.

Consider, as a field theory illustration of the instability,
\begin{equation}
{\cal{I}}=F_{\mu\nu}F^{\mu\nu} \ ,
\label{sourceMax}
\end{equation}
where $F=dA$ is the Maxwell tensor, in Minkowski spacetime. The scalar-free configuration is the Coulomb solution (in standard spherical coordinates):
\begin{equation}
\phi=0 \ , \qquad A=\frac{Q}{r}dt \ . 
\label{Coulomb}
\end{equation}
Consider also (in appropriate units) $f(\phi)=(1-\phi^2)^{-1}$. These choices obey $f'(0)=0$ and $\mu_{\rm eff}^2<0$. Thus~\eqref{Coulomb} is unstable against scalarisation. This model admits a simple \textit{closed form} spherically symmetric scalarised solution: 
\begin{equation}
\phi=\zeta \sin \left(\frac{Q}{r}\right),~
A=\left[\left(1-\frac{\zeta ^2}{2}\right)\frac{Q}{r}
+\frac{\zeta ^2}{4}\sin \left(\frac{2Q}{r}\right)\right]dt \ ,
\label{emgscalarised}
\end{equation}
where the integration constant obeys $|\zeta|<1$, to avoid divergences in the coupling. The total energy of the configurations is $E=4\pi \int_{r_0}^\infty \rho dr$, where $\rho=-T^t_t$ is the energy density. We consider a conducting sphere at $r=r_0$ and solutions~\eqref{Coulomb}-\eqref{emgscalarised} only for $r\geqslant r_0$, to regularise the total energy. The energies of these exterior configurations are, respectively, $E^{(\phi=0)}$ and $E^{(\phi \neq 0)}$. 
Then, $E^{(\phi \neq 0)}-  E^{(\phi=0)}= \pi {\zeta^2Q} \sin\left(\frac{2Q}{r_0}\right)$.
The scalarised solution is energetically favoured in a set of \textit{bands}, defined as ${Q}/r_0\in \, \pi ]n+1/2,n+1[$. The integer $n\in \mathbb{N}_0$, labelling the bands also counts, via~\eqref{emgscalarised}, the number of nodes exterior to $r_0$ of the scalar field profile. Thus, within these bands, the instability (likely) evolves dynamically~\eqref{Coulomb} into~\eqref{emgscalarised}.

\noindent{\bf {\em Spontaneous scalarisation of BHs.}} 
The toy model shows that spontaneous scalarisation $(i)$ is not exclusive of gravitational models and $(ii)$ can be supported by an electromagnetic non-minimal coupling. We shall now focus on the case of BHs and consider the gravitational model ($G=1=c$):
\begin{equation}
\mathcal{S}=\frac{1}{16\pi}\int d^4x\sqrt{-g}R+\mathcal{S}_\phi \ ,
\label{generalaction}
\end{equation}
where $R$ is the Ricci scalar. On a spherical, scalar-free BH solution,
with a generic line element
\begin{equation}
ds^2=-N(r)e^{-2\delta(r)}dt^2+\frac{dr^2}{N(r)}+r^2(d\theta^2+\sin^2 \theta d\varphi^2) \ , \ 
\label{metric}
\end{equation}
performing a (real) spherical harmonics decomposition of the scalar field
$\phi(r,\theta,\varphi)=\sum_{\ell m} Y_{\ell m}(\theta,\varphi)U_{\ell}(r)$,
the scalar field equation becomes
\begin{eqnarray}
\label{eqf1}
\frac{e^{\delta}}{r^2 }\frac{d}{dr}\left( \frac{r^2 N}{e^\delta} \frac{dU_{\ell}}{dr} \right)
-
\left[
      \frac{\ell(\ell+1)}{r^2}+\mu_{\rm eff}^2
			\right]
			        U_{\ell}=0 \ .
\end{eqnarray}
Eq. (\ref{eqf1}) is an eigenvalue problem:
for a given $\ell$, requiring an asymptotically vanishing, smooth
scalar field selects a discrete set of BHs. These are the \textit{bifurcation points} of the scalar-free solution. The (test) scalar field profiles they support - hereafter scalar \textit{clouds} - are distinguished by the node number of $U_\ell(r)$, $n\in \mathbb{N}_0$, besides $(\ell,m)$.

In~\cite{Doneva:2017bvd,Silva:2017uqg,Antoniou:2017acq} the model~\eqref{generalaction} was studied with ${\cal I}=\mathcal{L}_{GB}$, where $\mathcal{L}_{GB}$ is the Gauss-Bonnet invariant, and various different coupling functions, satisfying $f'(0)=0$ and $\mu_{\rm eff}^2<0$ (see also \cite{Cardoso:2013fwa,Cardoso:2013opa}). In this case the spontaneous scalarisation of the Schwarzschild solution is induced by this higher curvature correction, which is non-trivial for those BHs. Here, we shall take the source~\eqref{sourceMax}. Then, scalarisation of the Reissner-Nordstr\"om (RN) BH is induced without the need of higher curvature corrections. 
%

\noindent{\bf {\em Scalarisation in EMS models.}} 
We are interested in models~\eqref{generalaction} with~\eqref{actionS} and \eqref{sourceMax} which: $(i)$ admit 
the scalar-free RN solution. This rules out the usual Einstein-Maxwell-dilaton model~\cite{Garfinkle:1990qj}, where  $ f(\phi)=e^{-\alpha \phi}$; $(ii)$
approach the standard Einstein-Maxwell system in the far field, $i.e.$ $\phi\to 0$ and $f(\phi)\rightarrow 1$ as $r\to \infty$.

From the scalar equation of motion, it is possible to derive two Bekenstein type identities~\cite{Bekenstein:1972ny} that set the following constraints on $f$ (for a purely electric field $F^2<0$): $f_{,\phi \phi} >0$ and $\phi f_{,\phi}>0$ for some range of the radial coordinate $r$.
A simple potential
compatible with the above requirements, that we shall use hereafter, is
\begin{eqnarray}
\label{f}
 f(\phi)=e^{-\alpha \phi^2}.
\end{eqnarray}  
The coupling $\alpha$
is a dimensionless constant and the conditions on $f$ imply  $\alpha<0$
  for  $F^2<0$. 

The RN (scalar-free) solution is given by~\eqref{Coulomb} and~\eqref{metric}, with $\delta=0$, $N(r)=1-2M/r+Q^2/r^2$. The scalar perturbations on this background are given by~\eqref{eqf1} with $\mu^2_{\rm eff}=\alpha Q^2/r^4<0$, hence exhibiting the instability. For $\ell=0$, one finds an \textit{exact} test field solution
\begin{eqnarray}
\label{ex1}
U_0(r)=P_u 
\left[
1+\frac{2Q^2(r-r_H)}{r(r_H^2-Q^2)}
\right],
\end{eqnarray}
where $u\equiv(\sqrt{4\alpha+1}-1)/2$, $r_H\equiv M+\sqrt{M^2-Q^2}$ and $P_u$ is a Legendre function.
For generic
parameters  $(\alpha,Q,r_H)$, finding the $\ell=0$ bifurcation points from RN reduces to studying the zeros of this function as  $r\to \infty$. Examples are shown in Fig.~\ref{profile2} below. Bifurcation requires $\alpha$ below a maximal value, $\alpha_{\rm max}$. For fundamental modes ($n=0$) we obtain 
$\alpha_{\rm max}=-1/4$ $(\ell=0)$,
$\alpha_{\rm max}\simeq -2.784$ $(\ell=1)$
and
$\alpha_{\rm max}\simeq -7.087$ $(\ell=2)$.

\begin{figure}[h!]
\begin{center}
\includegraphics[width=0.49\textwidth]{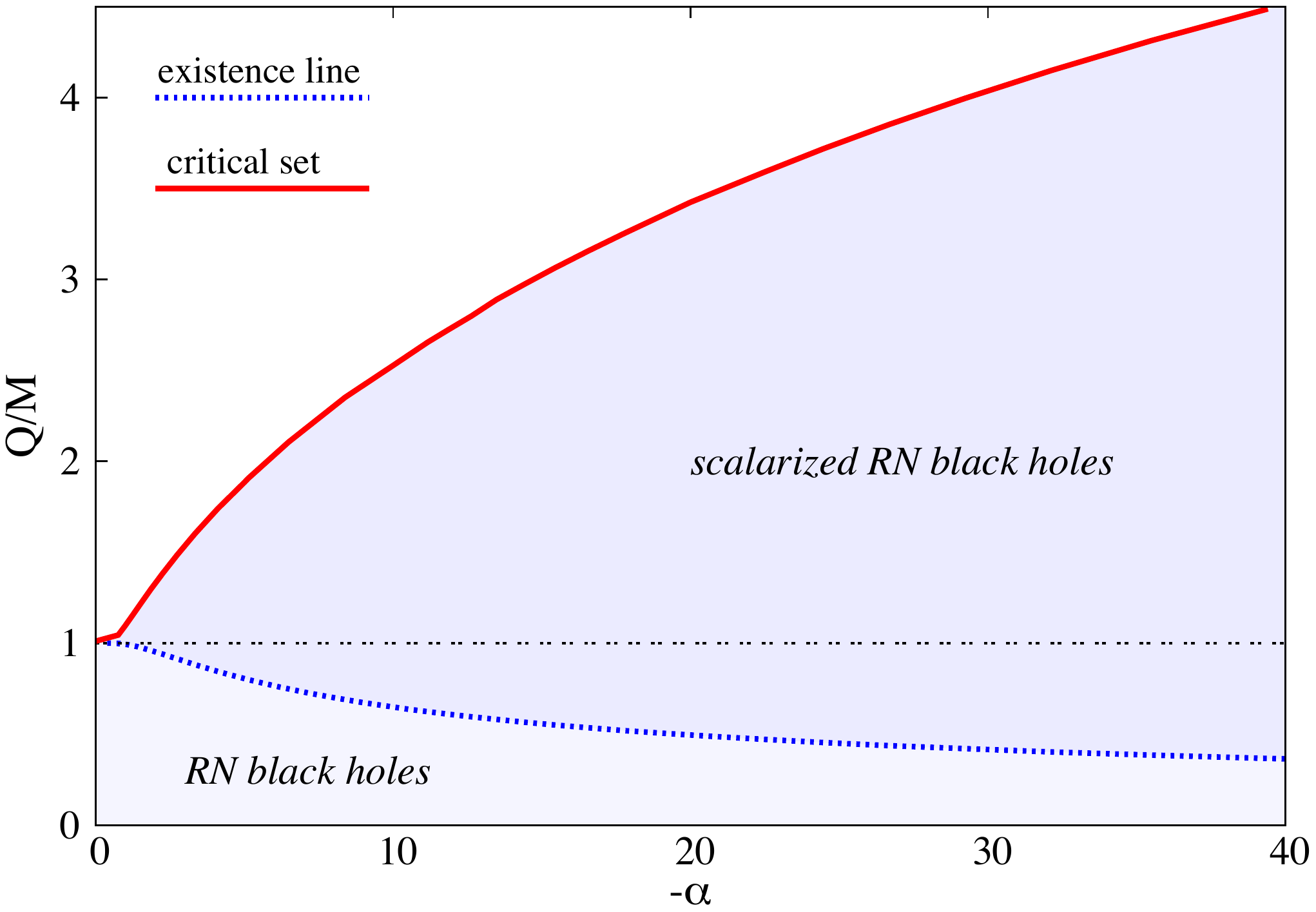}  
\caption{ 
Part of the domain of existence of spherical EMS scalarised BHs with $n=0$ in the $(\alpha,q)$-plane. 
}
\label{domain}
\end{center}
\end{figure}

\noindent{\bf {\em \textit{Spherical sector (no nodes): domain of existence.}}} 
The spherical scalarised BHs bifurcate from the RN solution for any $\alpha<-1/4$. They are the non-linear realisations of the $(n,\ell,m)=(0,0,0)$ clouds. These non-linear solutions were obtained by using the ansatz~\eqref{metric}, $\phi=\phi(r)$, $A=V(r)dt$, and standard numerical techniques to solve coupled non-linear ODEs  - see Appendix~\ref{apA} for details. Varying $\alpha$ we have obtained the domain of existence of spherical scalarised BHs (with $n=0$) shown in Fig.~\ref{domain}.

For each $\alpha$, a branch of scalarised solutions bifurcates from a RN BH with a particular charge to mass ratio $q\equiv Q/M$. As $\alpha$ varies, these RN BHs define an~\textit{existence line}. Each constant $\alpha$-branch ends at a \textit{critical}, (likely) singular, configuration: the numerics indicate the Kretschmann scalar and the horizon temperature diverge, the horizon area vanishes, whereas the mass and the scalar ``charge" (defined as $Q_s\equiv - \lim_{r\rightarrow \infty} r^2d\phi/dr$) remain finite. Along the $\alpha$=\textit{constant} branch, $q$ increases beyond unity. Thus, scalarised BHs can be \textit{overcharged}.

\begin{figure}[h!]
\begin{center}
\includegraphics[width=0.49\textwidth]{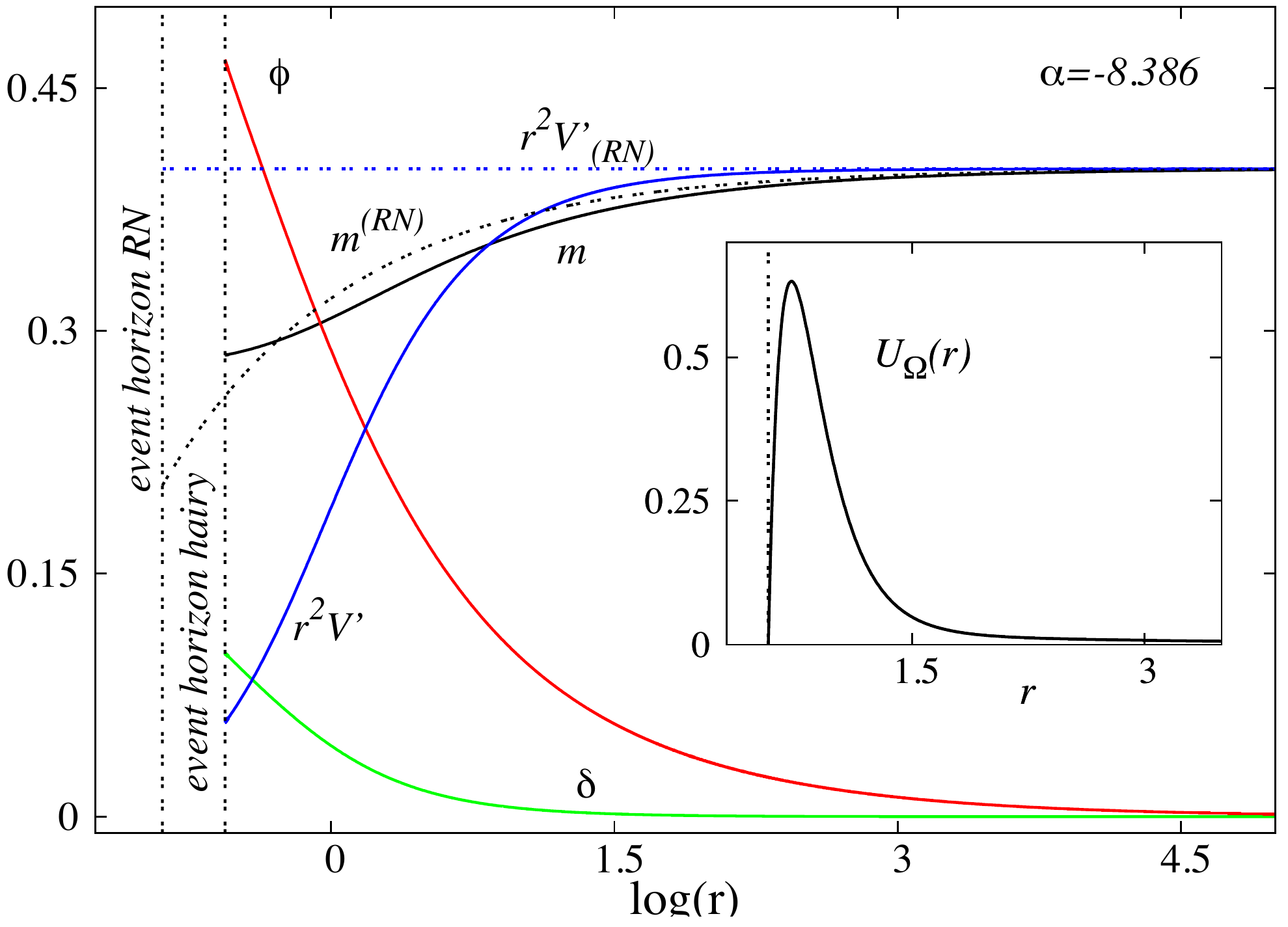}  
\caption{ 
Profiles for a typical spherical scalarizsed BH  (solid lines) and the RN BH (dotted lines) with the same $M = 0.4002$ and $Q = 0.4$.  The horizon areal radius is smaller for the RN BH. 
The inset shows the corresponding potential $U_\Omega$.
}
\label{profile1}
\end{center}
\end{figure}

The  domain of existence of scalarised BHs is bounded by:
$(i)$ the existence line and
$(ii)$ the set of all critical solutions. 
In a part of the domain of existence there is non-uniqueness: RN and scalarised solutions co-existe with the same global charges - Figs.~\ref{domain} and~\ref{profile1}. In this region, the scalarised solutions are always entropically favoured, $cf.$ Fig.~\ref{profile2}. These  spherical scalarised BHs are candidate endpoints of the \textit{spherical} evolution (if adiabatic) of the linearly unstable RN BHs in the EMS model. Further evidence  is provided by computing spherical perturbations, with frequency $\Omega$, of the scalarised solutions. The problem reduces to a single 1D Schr\"odinger-like equation with the potential
$
U_\Omega=\frac{N}{e^{2\delta} r^2}
\left[
\frac{e^{\alpha \phi^2}Q^2[2(\alpha \phi+r\phi')^2+ \alpha-1]}{r^2}
+1-N-2r^2 \phi'^2
\right]
$.
This potential vanishes both at the BH event horizon  and at infinity, being regular everywhere in between.
It follows that the Schr\"odinger eq. will have no bound
states if the potential is everywhere positive.
For all $n=0$ solutions analysed, this positivity is indeed satisfied. The absence of such bound states guarantees the stability of the scalarised BHs against this class of perturbations.

\begin{figure}[h!]
\begin{center}
\includegraphics[width=0.49\textwidth]{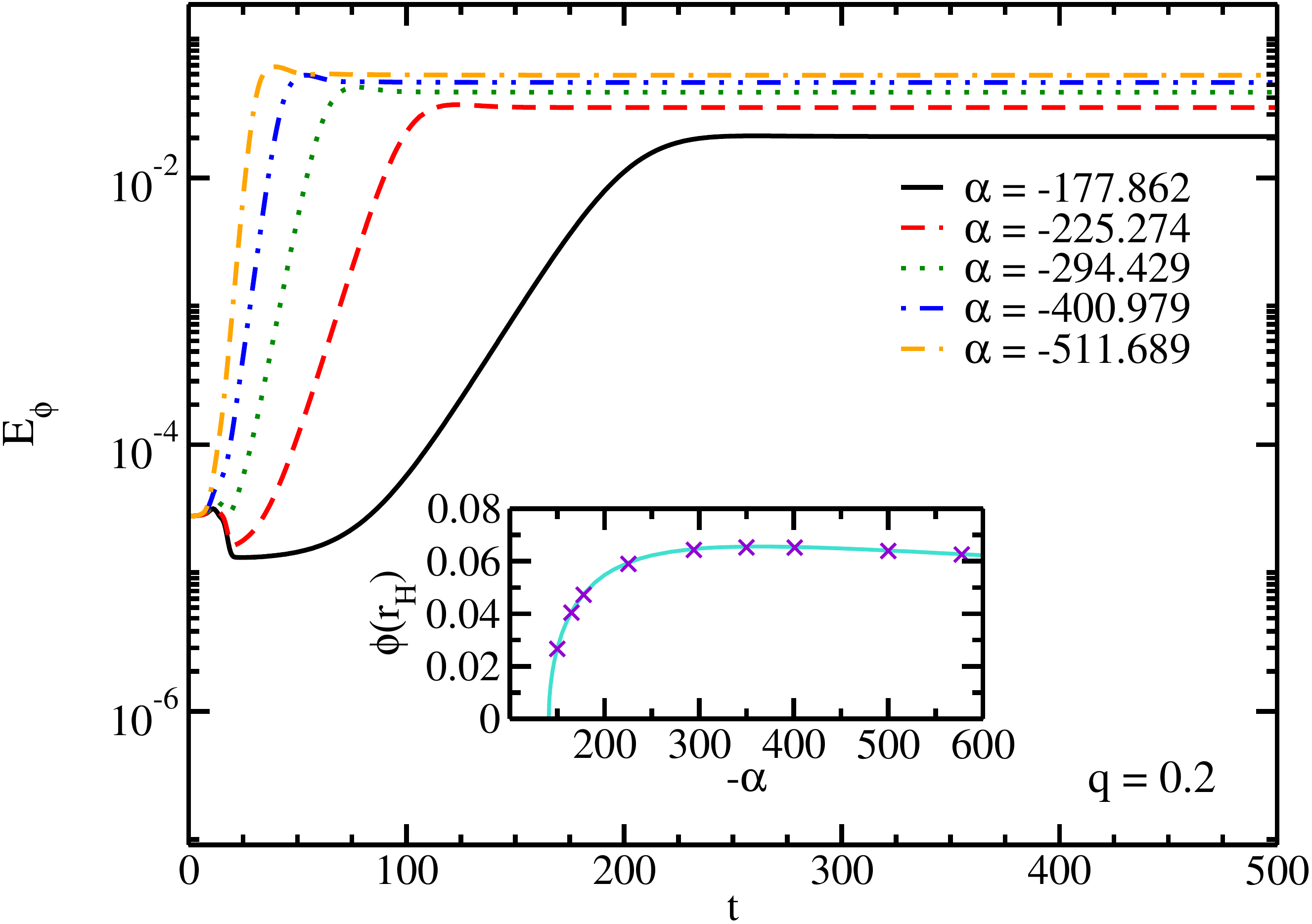}  
\caption{ 
Evolution of  $E_{\phi}$ outside an unstable RN BH with $q=0.2$, for different couplings. (Inset) Scalar field value at the horizon radius:  end state of the evolutions (crosses) and static solutions (line) with $q=0.2$.
}
\label{evolution}
\end{center}
\end{figure}

A final piece of \textit{dynamical} evidence that the scalarisation of the RN BH leads to the solutions in Fig.~\ref{domain} comes from performing fully non-linear numerical evolutions - see~\cite{Sanchis-Gual:2015lje} and Appendix~\ref{apB}  for details. We start with a RN BH and a small Gaussian perturbation for $\phi$ and monitor the scalar field energy,  $E_\phi=\int_{r_H}^\infty n^{\alpha}n^{\beta}T^{\rm{SF}}_{\alpha\beta} dV$, where $T^{\rm{SF}}$ is the $\phi$ stress-energy tensor and $n$ the  4-velocity of the Eulerian observer in the 3+1 spacetime decomposition~\cite{Alcubierre08a}. Fig.~\ref{evolution} shows the time evolution of $E_\phi$ for an initial RN BH with $q=0.2$ and different couplings.
One observes an initial growth of $E_\phi$ followed by a saturation and equilibrium. For the largest $|\alpha|$ saturation is faster and it is preceded by an overshooting of the equilibrium value. The inset shows the value of the scalar field at the horizon (areal) radius, both after equilibrium is reached in each simulation (crosses) and from the scalarised BHs with the same $q=0.2$ computed as static solutions (line). The values agree within 1\%. Thus, the end point of the evolutions are the perturbatively stable scalarised BHs with the same $q$ as the initial unstable RN BH. 
A larger set of numerical evolutions, exploring the $(q,\alpha)$ space, confirms  scalarisation is generic and leads to the BHs in Fig.~\ref{domain}, but for larger $q$ values of the initial RN BH, the evolution does not preserve (even approximately) $q$. A discussion of these evolutions will be presented elsewhere.

\noindent{\bf {\em \textit{Non-spherical sector: multipole freedom.}}} 
Generic $(n,\ell,m)$-scalar clouds also have non-linear realisations, yielding branches of non-spherical scalarised BHs bifurcating from the RN trunk.  We have constructed such solutions using the Einstein-De Turck approach~\cite{Headrick:2009pv,Adam:2011dn} - see~\cite{Dias:2015nua} for a review and  Appendix~\ref{apC} for details. 

All configurations constructed  are regular on and outside a topologically (but not geometrically) spherical horizon and asymptotically flat. Solutions bifurcating from a scalar cloud with $m=0$ ($m\neq 0$) are (are not) axially symmetric. The latter, in fact, have no spatial isometries. 
They
provide the first explicit example of 
static, asymptotically flat BHs without any continuous (spatial) symmetries (see~\cite{Ridgway:1995ke,Ioannidou:2006mg,Herdeiro:2016plq} for related work).

In Fig.~\ref{profile2} we exhibit results concerning solutions with $(n,\ell,m)=(0,1,0); (0,1,1); (0,2,0); (0,2,1); (0,2,2)$ and  $(n,0,0)$, for an illustrative value of the coupling $\alpha$.  The latter are for $n=1,2,3$, corresponding to excited spherical scalarised BHs. The top panel shows that as either $\ell$ or $n$ increases, the bifurcation point moves towards extremality of the scalar-free RN solution. This bifurcation point does not depend on $m$ (Fig.~\ref{profile2}, bottom panel) as anticipated from the linear theory analysis. Also, the relative location of the bifurcation point for the first few values of $\ell,n$ can be seen in the inset of the bottom panel. We remark that when continuing scalar clouds with $\ell,m\neq 0$ into the non-linear regime, the corresponding scalar field does not remain a pure $\ell,m$ mode; non-linearities excite all $\ell$ modes with the same $m$ and parity.

 \begin{figure}[h!]
\begin{center}
\includegraphics[width=0.49\textwidth]{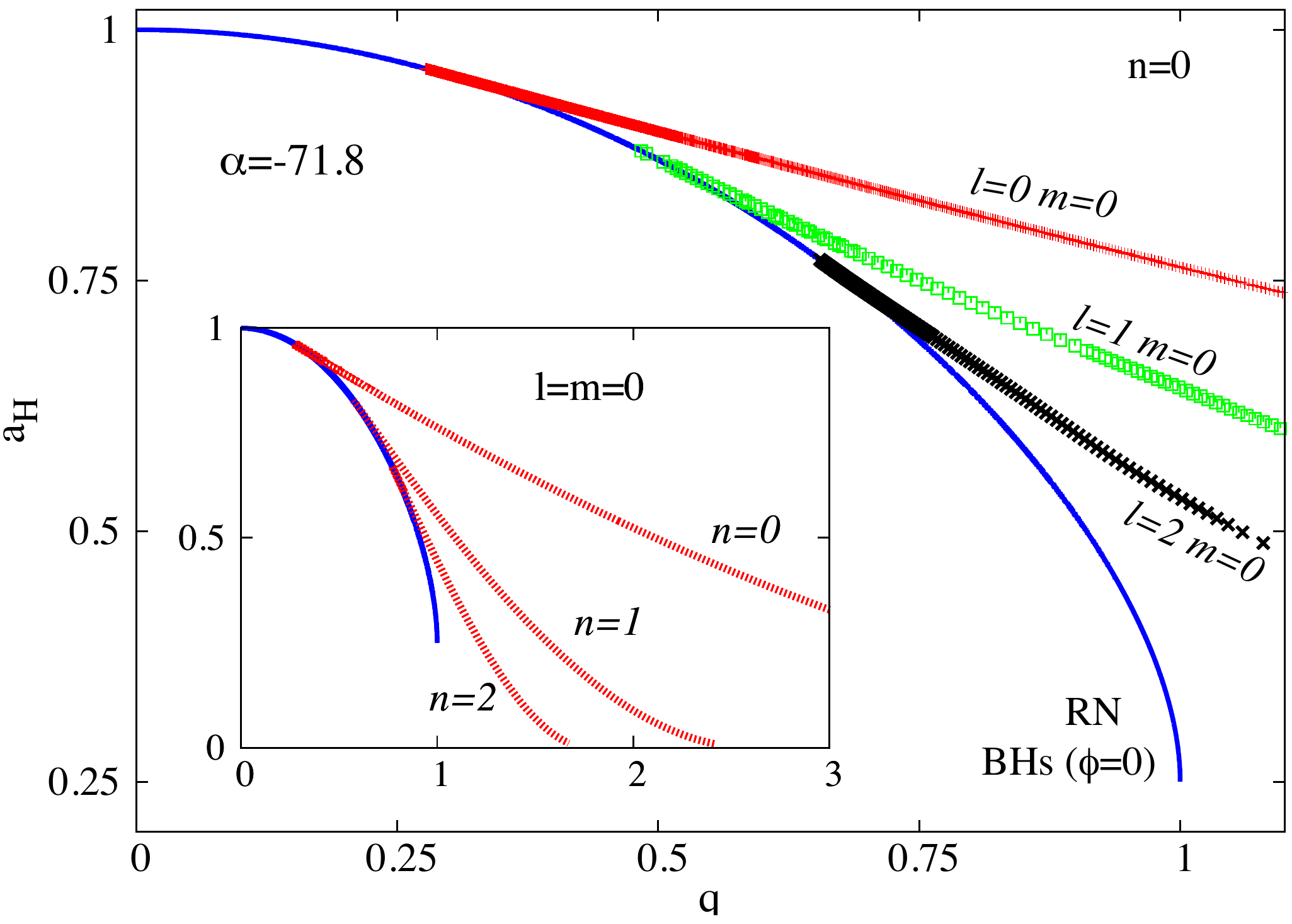} 
\includegraphics[width=0.49\textwidth]{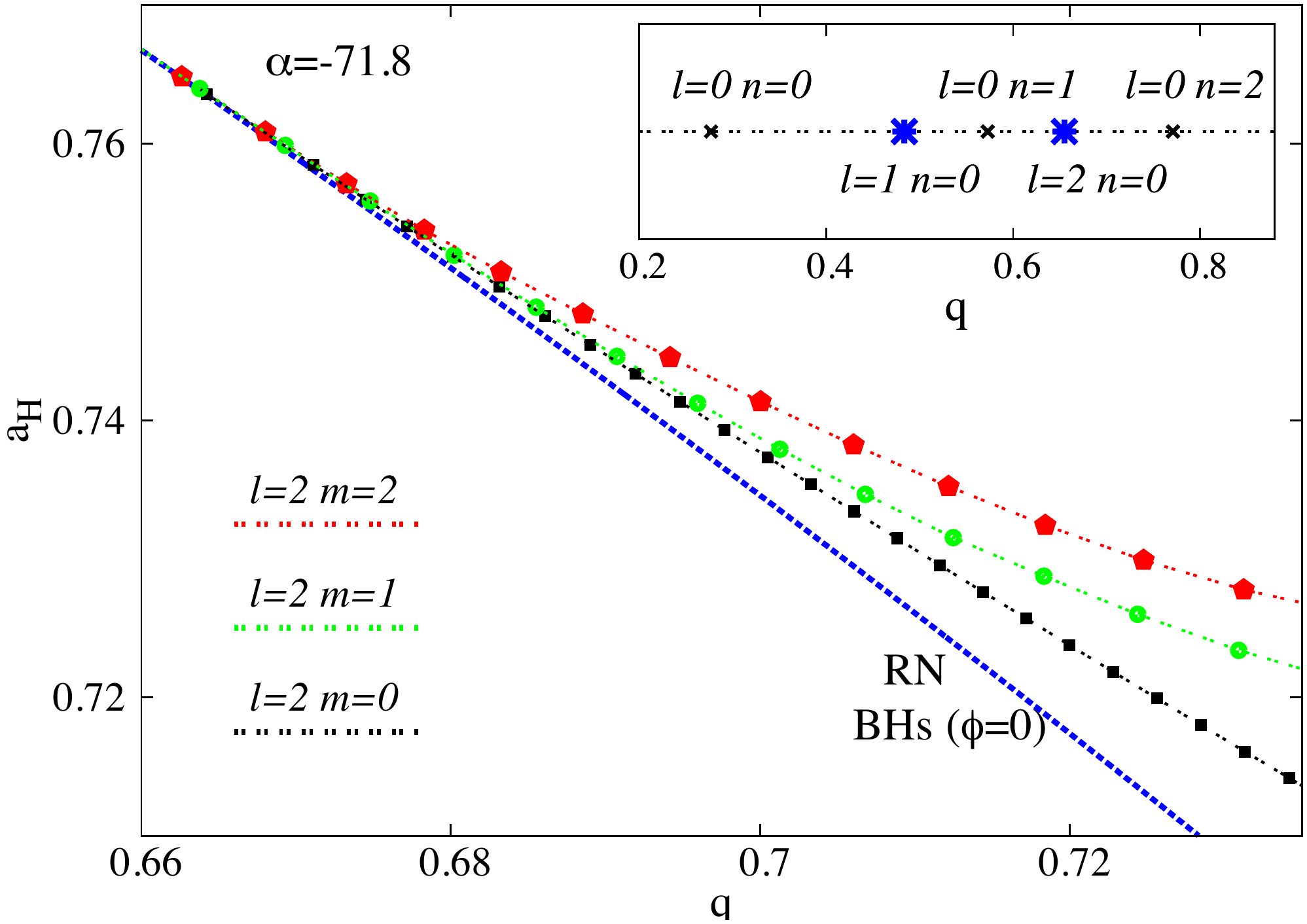}
\caption{ 
Scalarised BHs 
with $\alpha=-71.8$, in a $q\equiv Q/M$ $vs.$ $a_H\equiv \frac{A_H}{16\pi M^2}$ diagram, where $A_H$ is the horizon area. 
BHs with $m\neq 0$ have only discrete spatial symmetries.
}
\label{profile2}
\end{center}
\end{figure} 

All scalarised solutions are entropically preferred over a comparable RN solution (with the same total charge and mass). Within the scalarised solutions, the preferred one is the fundamental state with $\ell=0=m$, $i.e.$ spherically symmetric. Also, for all studied cases, for constant $m$, the entropy is maximised by the branch of solutions emerging from the $\ell=m$ zero mode. These entropic arguments are, however, inconclusive for dynamical considerations.

\noindent{\bf {\em \textit{General dynamics: the charged drop analogy.}}} 
 Consider a \textit{generic} ($i.e.$ non-spherical) perturbation of the RN BH with sufficiently large $q$. The existence of the  non-spherical branches of scalarised BHs allows non-spherical perturbations to \textit{grow} rather than damp (as they would in electrovacuum). A suggestive parallelism with charged drops in fluid dynamics can be drawn. 

A self-gravitating, uncharged, isolated liquid mass is spherical and stable against small perturbations~\cite{doi:10.1112/plms/s1-11.1.57}. But if the liquid is electrically charged, conducting and surrounded by an insulator ($e.g.$ a gas), the competition between the cohesive tension and the electric repulsion makes the spherical drop (which remains a solution)  \textit{unstable}, for charge beyond the Rayleigh limit~\cite{doi:10.1080/14786448208628425}. New branches  of non-spherical solutions of the fluid-electrostatic equations, associated to each spherical harmonic, emerge, bifurcating from the trunk family of spherical fluid balls. Such structure is analogous to that shown in Fig.~\ref{profile2} - see, $e.g.$ Fig. 2 in~\cite{doi:10.1063/1.857551}. Both experiments~\cite{nature} and simulations~\cite{PhysRevLett.106.144501} show that the unstable charged drop evolves towards a non-spherical shape.

BHs in the EMS model (unlike charged drops) have a $\ell=0$ mode instability which may always dominate and lead to a spherical scalarised BHs. If, however, there are initial conditions such that $\ell\neq0$ modes dominate, a dynamical symmetry breaking may occur. Fully non-linear numerical evolutions in the EMS model with current technology can probe this possibility.

\bigskip

\noindent{\bf {\em Acknowledgements.}}
C.H. would like to thank E. Berti for discussions. This work has been supported by the FCT (Portugal) IF programme, by the CIDMA (FCT) 
strategic project UID/MAT/04106/2013, by the Spanish MINECO (grant AYA2015-66899-C2-1-P), by the Generalitat Valenciana (PROMETEOII-2014-069, ACIF/2015/216) and by  the  European  Union's  Horizon  2020  research  and  innovation  (RISE) programmes H2020-MSCA-RISE-2015 Grant No.~StronGrHEP-690904 and H2020-MSCA-RISE-2017 Grant No.~FunFiCO-777740. The authors would like to acknowledge networking support by the
COST Action CA16104. Computations were performed at the Blafis cluster in Aveiro University and at the Servei d'Inform\`atica de la Universitat de Val\`encia.

\bigskip


\appendix

\section{Static spherical scalarised solutions}
\label{apA}
Consider the metric ansatz~\eqref{metric}, with $N(r)=1-2m(r)/{r}$, where $m(r)$ is the Misner-Sharp mass function~\cite{Misner:1964je}.
The
gauge connection ansatz is purely electric, $A=V(r) dt$ and the scalar field is a function of $r$ only. This ansatz yields the following
effective Lagrangian:
\begin{eqnarray}
\label{Leff}
\mathcal{L}_{eff}=e^{-\delta}m'-\frac{1}{2}e^{-\delta}r^2 N \phi'^2+\frac{1}{2}e^{\delta-\alpha \phi^2}r^2 V'^2 \ ,
\end{eqnarray}
the corresponding equations of motion being
\begin{eqnarray}
\label{ec-m}
&&
m'=\frac{1}{2} r^2 N \phi'^2+\frac{1}{2}e^{2\delta-\alpha \phi^2}r^2 V'^2 \ ,\nonumber
\\
\label{ec-d}
&&
\delta'+r \phi'^2=0 \ , \nonumber 
\\
&& 
\label{ec-V}
( e^{\delta-\alpha \phi^2 }r^2 V')'=0 \ , \nonumber
\\
&&
\label{ec-phi}
 (e^{-\delta} r^2 N\phi')'=\alpha e^{\delta-\alpha \phi^2}\phi r^2 V'^2\ .
\end{eqnarray}
The electric potential  can be eliminated from the above equations noticing the existence of a first integral,
\begin{eqnarray}
\label{first-int}
V'= -e^{-\delta+\alpha \phi^2 } \frac{Q}{r^2}  \ ,
\end{eqnarray}
where $Q$ is the electric charge.

We assume the existence of a horizon located
at $r=r_H>0$. In its vicinity one finds the following 
approximate solution to the field equations
\begin{eqnarray}
\label{horizon1}
&&
m(r)=\frac{r_H}{2}+m_1(r-r_H)+\dots,~~ \nonumber  \\
&&
\delta(r)=\delta_0 + \delta_1 (r-r_H)+\dots,~~ \nonumber
\\
&&
\nonumber
\phi(r)=\phi_0 + \phi_1 (r-r_H)+\dots,~~ \\ &&
V(r)=v_1 (r-r_H)+\dots,~~
\end{eqnarray}
where
\begin{eqnarray}
\label{horizon2}
&&
m_1=\frac{e^{\alpha \phi_0^2}Q^2}{2r_H^2} , ~~
 \phi_1=\frac{1}{r_H} \frac{\alpha \phi_0 Q^2  e^{\alpha \phi_0^2}}{r_H^2-Q^2 e^{\alpha\phi_0^2}} \ ,~~ \nonumber  \\ &&
\delta_1=-\phi_1^2 r_H\ ,~~
 v_1=-\frac{e^{\alpha \phi_0^2-\delta_0}Q}{ r_H^2} \ .~~
\end{eqnarray}
This approximate solution contains two essential
parameters,
$\phi_0$
and 
$\delta_0$,  
which are found matching the above expansion with
the following asymptotics of the solutions in the far field:
\begin{eqnarray}
\label{inf1}
&& m(r)=M-\frac{Q^2+Q_s^2}{2r}+\dots \ , 
~~\phi(r)=\frac{Q_s}{r}+\dots \ ,~~ \nonumber \\
&&
V(r)=\Phi+\frac{Q}{r}+\dots,~~
\delta(r)=\frac{Q_s^2}{2r^2}+\dots.
\end{eqnarray}
Apart from $M$ and $Q$, the essential parameters here are
$\Phi$ (the electrostatic potential)
and
$Q_s$
(the scalar ``charge").

The solutions satisfy the
virial identity:
\begin{eqnarray}
\label{v1} 
&& \int_{r_H}^\infty dr \left\{
e^{-\delta}  \phi'^2
  \left[
1+\frac{2r_H}{r}\left(\frac{m}{r}-1\right) 
   \right]
\right\} \nonumber  \\
&&
=
\int_{r_H}^\infty dr \left [
e^{-\delta+\alpha \phi^2}  
\left(1-\frac{2r_H}{r}\right)\frac{Q^2}{r^2}
\right ] \ .
\end{eqnarray}
Since the term in square brackets, in the first line is always non-negative, this shows that a nontrivial scalar field
can only be supported if $Q\neq 0$.

The data at infinity is specified by the ADM mass $M$,
electric charge $Q$, electrostatic potential $\Phi$
and $Q_s$.
The  horizon data  is the Hawking temperature and horizon area $
T_H=\frac{1}{4\pi}N'(r_H)e^{-\delta(r_H)}$, $A_H=4\pi r_H^2$,
together with $\phi(r_H)\equiv \phi_0$.
Interestingly, one can show that the Smarr relation 
is not affected by the scalar hair, $M=\frac{1}{2} T_H A_H+\Phi Q$, the first law of BH thermodynamics being $dM=\frac{1}{4}T_H dA_H+\Phi dQ$. 

We have noticed that the solutions satisfy the following interesting relation
\begin{eqnarray}
\label{figen}
 M^2+Q_s^2=Q^2+\frac{1}{4}A_H^2 T_H^2\ .
\end{eqnarray}
Remarkably, one can shown that 
(\ref{figen})
holds for any choice of the coupling function 
$f(\phi)$
(with $\phi\to Q_s/r$ as $r\to \infty$).

One observes that the solutions do not change when considering the scaling $
r\to \lambda r,~~Q\to \lambda Q$,
while $\alpha$ is fixed. $q,a_H$ are invariant under these transformations.

\section{Numerical evolutions}
\label{apB}
To address numerical evolutions in the EMS system, we have modified the code described in~\cite{Sanchis-Gual:2015lje, Sanchis-Gual:2016tcm} and adapted the evolution equations in~\cite{Hirschmann:2017psw} to the coupling described above. The 3+1 metric split reads $ds^2=-(\alpha_0^2+\beta^r \beta_r)dt^2+2\beta_r dtdr+e^{4\chi}\left[a\, dr^2+ b\, r^2 d\Omega_2\right]$, where the lapse $\alpha_0$, shift component $\beta^r$, and the (spatial) metric functions, $\chi,a,b$ depend on $t,r$. The electric field $E^\mu=F^{\mu\nu} n_\nu$ has only a radial component and the magnetic field $B^\mu=\star F^{\mu\nu} n_\nu$ vanishes, where $n^\mu$ is the 4-velocity of the Eulerian observer~\cite{Torres:2014fga}. 

  As in~\cite{Sanchis-Gual:2015lje}, we take as initial state a RN BH with ADM mass $M$, and charge $Q$, together with a vanishing scalar field. We set, as the scalar field initial data, a  Gaussian distribution of the form 
  \begin{equation} \phi=A_0e^{-(r-r_0)^2/\lambda^2} \ ,
  \end{equation} 
  with  $A_0=3\times 10^{-4}$, $r_0=10M$ and $\lambda=\sqrt{8}$. As the geometrical initial data, we choose a conformally flat metric with $a=b=1$ together with a time symmetry condition, $i.e.$ vanishing extrinsic curvature, $K_{ij}=0$. 
The conformal factor is given by
\begin{equation}
\psi  \equiv  e^{ \chi}= \biggl[\biggl(1+\frac{M}{2r}\biggl)^{2}-\frac{Q^{2}}{4r^{2}}\biggl]^{1/2}.
\end{equation}\\

At $t=0$, we choose a ``pre-collapsed" lapse $\alpha_0 = \psi^{-2}$ and a vanishing shift $\beta^{r}=0$. Initially, the electric field is given by $E^{r}=\frac{Q}{r^{2}\psi^{6}}$.

 The logarithmic numerical grid extends from the origin to $r=10^3M$ and uses a maximum resolution of
$\Delta r=0.0125M$. A broader survey of the parameter space will be presented elsewhere.


In spherical symmetry, the evolution equations for the electric field and an extra variable, $\Psi$, to damp dynamically the constrains, take the form
\begin{eqnarray}
\partial_{t}E^{r}&=&\beta^{r}\partial_{r}E^{r}-E^{r}\partial_{r}\beta^{r}+(\alpha_0 K E^{r} -D^{r}\Psi)\nonumber\\
&&-2\alpha\alpha_0\,\phi\Pi E^{r}\,, \nonumber \\
\partial_{t}\Psi&=&\alpha_0(2\alpha \phi D_{r}\phi E^{r} - D_{i}E^{i}-\kappa_{1}\Psi)\,,
\end{eqnarray}
where $K$ is the trace of  $K_{ij}$, we have taken $\kappa_1=1$ and $\Pi\equiv -n^{a}\nabla_{a}\phi$.

To solve the Klein-Gordon equation we evolve the following system of first-order equations: 
\begin{eqnarray}
\partial_{t}\phi&=&\beta^{r}\partial_{r}\phi-\alpha_0\Pi \  , \nonumber \\
\partial_{t}\Pi&=&\beta^{r}\partial_{r}\Pi+\alpha K\Pi-\frac{\alpha_0}{ae^{4\chi}}\biggl[\partial_{rr}\phi\nonumber\\
&+&\partial_{r}\phi\biggl(\frac{2}{r}-\frac{\partial_{r}a}{2a}+\frac{\partial_{r}b}{b}+2\partial_{r}\chi\biggl)\biggl]\nonumber\\
&-&\frac{\partial_{r}\phi}{ae^{4\chi}}\,\partial_{r}\alpha_0+\alpha\alpha_0\phi\, e^{-\alpha\Phi^{2}}\,a\,e^{4\chi}(E^{r})^{2} \ .
\label{eq:sist-KG}
\end{eqnarray}


The matter source terms  for the scalar field, to be used in the Einstein equations, read
\begin{eqnarray}
\mathcal{E}^{\rm{SF}}&\equiv &n^{\alpha}n^{\beta}T^{\rm{SF}}_{\alpha\beta}=\frac{1}{8\pi}\biggl(\Pi^{2}+\frac{\partial_{r}\phi^{2}}{ae^{4\chi}}\biggl) \nonumber \\
j_{r}^{\rm{SF}}&\equiv &-\gamma^{\alpha}_{r}n^{\beta}T^{\rm{SF}}_{\alpha\beta}=-\frac{1}{4\pi}\Pi\,\partial_{r}\phi\ ,\nonumber \\
S_{a}^{\rm{SF}}&\equiv &(T^{r}_{r})^{\rm{SF}}=\frac{1}{8\pi}\biggl(\Pi^{2}+\frac{\partial_{r}\phi^{2}}{ae^{4\chi}}\biggl)\ , \nonumber \\
S_{b}^{\rm{SF}}&\equiv &(T^{\theta}_{\theta})^{\rm{SF}}=\frac{1}{8\pi}\biggl(\Pi^{2}-\frac{\partial_{r}\phi^{2}}{ae^{4\chi}}\biggl)\ .
\end{eqnarray}
and for the electric field
\begin{eqnarray}
\mathcal{E}^{\rm{em}}=-S_{a}^{\rm{em}}=S_{b}^{\rm{em}}=\frac{1}{8\pi}\,a\,e^{4\chi}(E^{r})^{2}e^{-\alpha\phi^{2}}\ .
\end{eqnarray}

The momentum density $j_{r}^{\rm{em}}$ vanishes because there is no magnetic field in spherical symmetry.

\section{Static non-spherical scalarised solutions}
\label{apC}
In the Einstein-De Turck (EDT) approach  one solves the so called Einstein-DeTurck equations 
\begin{eqnarray}
\label{EDT}
R_{\mu\nu}-\nabla_{(\mu}\xi_{\nu)}= 2 \left(T_{\mu\nu}-\frac{1}{2}T  g_{\mu\nu}\right) \ .
\end{eqnarray}
$\xi^\mu$ is a vector defined as
$
\xi^\mu\equiv g^{\nu\rho}(\Gamma_{\nu\rho}^\mu-\bar \Gamma_{\nu\rho}^\mu)\ ,
$
where 
$\Gamma_{\nu\rho}^\mu$ is the Levi-Civita connection associated to the
spacetime metric $g$ that one wants to determine, and a reference metric $\bar g$ is introduced, 
($\bar \Gamma_{\nu\rho}^\mu$ being the corresponding Levi-Civita connection).
Solutions to (\ref{EDT}) solve the Einstein equations
iff $\xi^\mu \equiv 0$ everywhere.
To achieve this,
we impose boundary conditions  which are compatible with
$\xi^\mu = 0$
on the boundary of the domain of integration.
Then, this should imply $\xi^\mu \equiv 0$ everywhere,
a condition which is verified from  the numerical output.

We use  a metric ansatz
with seven functions, $F_1,F_2,F_3,F_0,S_1,S_2,S_3$
which depend on $(r,\theta,\varphi)$:
\begin{eqnarray}
\label{metric2}
&&
ds^2=-F_0(r,\theta,\varphi) N(r)dt^2+F_1(r,\theta,\varphi) \frac{dr^2}{N(r)} 
\\
\nonumber
&&
+F_2(r,\theta,\varphi)\left(r d\theta+S_1(r,\theta,\varphi) dr \right)^2 \\
&&
+F_3(r,\theta,\varphi)  \big(r \sin \theta d\varphi+S_2(r,\theta,\varphi) dr+S_3(r,\theta,\varphi) r d\theta \big)^2  \ . \nonumber
\end{eqnarray}
The reference metric is that of the RN BH, with
$D_1=F_2=F_3=F_0=1$,
$S_1=S_2=S_3=0$
and
 $N(r)$ 
\begin{eqnarray}
\label{N}
N(r)=\left(1-\frac{r_H}{r}\right)\left(1-\frac{\bar{q}^2}{r r_H}\right),
\end{eqnarray}
where $r_H\geqslant 0$ is  the event horizon radius 
and $\bar{q}<r_H$ another input constant.

The ansatz for the Maxwell field is $
A=V(r,\theta,\varphi) dt$; for the scalar field it has the most general expression compatible with a static spacetime, $\phi=\phi(r,\theta,\varphi)$. 
The solutions have a nodeless scalar field $n=0$,
being indexed by two integers $(\ell,m)$,
which classify the  
corresponding  scalar zero mode.  
The particular case of the  axially symmetric  (or even spherically symmetric) BHs
 can also be studied within this framework for $m=0$.
For example, the  axially symmetric metric has $S_2=S_3=0$, and all remaining functions depend on $(r,\theta)$ only.

In this approach, the problem reduces to
solving a set of nine PDEs with suitable
boundary conditions (BCs).
The BCs are found by 
 constructing an approximate form of the solutions on the
boundary of the domain of integration compatible with the requirement $\xi^\mu = 0 $,
regularity  of the solutions and asymptotic flatness.

In the non-axisymmetric cases, we have studied $m>0$ solutions with a reflection symmetry along the equatorial plane
($\theta=\pi/2$) and two $\mathbb{Z}_2$-symmetries w.r.t. the $\varphi-$coordinate.
Then the domain of integration  for the $(\theta,\varphi)$-coordinates 
is $[0,\pi/2]\times [0,\pi/2]$.

In numerics, we have imposes the following BCs. At infinity:
\begin{eqnarray}
\label{infinity}
&& F_1= F_2=  F_3= F_0=1\ ,
 \nonumber \\
&&
S_1=S_2= S_3=0\ ,~r^2 \partial_r V=-Q\ , ~\phi=0\ ,
\end{eqnarray}
 where $Q$ is an input parameter fixing the electric charge;
at $\theta=0$:
\begin{eqnarray}
\label{t0}
&&
\partial_\theta F_1= \partial_\theta F_2=  \partial_\theta F_3= \partial_\theta F_0=0\ ,~
\nonumber \\ 
&&
S_1=S_2= \partial_\theta S_3=0\ , ~\partial_\theta V=0\ ,~
\end{eqnarray}
together with $\partial_\theta \phi=0$ for axially symmmetric BHs
($m=0$) and  $ \phi=0$ in rest;
at $\theta=\pi/2$:
\begin{eqnarray}
\label{tpi2}
&& \partial_\theta F_1= \partial_\theta F_2=  \partial_\theta F_3= \partial_\theta F_0=0\ ,~
  \nonumber \\
&&
S_1=\partial_\theta  S_2=   S_3=0\ ,~\partial_\theta  V =0\ ,
\end{eqnarray}
together with $\phi=0$,  except if
$\ell+m$ is an even number, in which case we impose  $\partial_\varphi \phi=0$;
at $\varphi=0$:
\begin{eqnarray}
\label{fi0}
&&\partial_\varphi F_1= \partial_\varphi F_2=  \partial_\varphi F_3= \partial_\varphi F_0=0\ ,~
 \nonumber \\
&&
\partial_\varphi S_1=   S_2=   S_3=0\ ,~\partial_\varphi  V= \partial_\varphi  \phi =0  \ ;
\end{eqnarray}
at
 $\varphi=\pi/2$:
\begin{eqnarray}
\label{fipi2}
&& \partial_\varphi F_1= \partial_\varphi F_2=  \partial_\varphi F_3= \partial_\varphi F_0=0\ ,~ \nonumber \\
&&
\partial_\varphi S_1=   S_2=   S_3=0\ ,~\partial_\varphi  V=0\ ,
\end{eqnarray}
together with
$ \phi=0$
for odd $m$
and
$\partial_\varphi  \phi=0$ for even $m$.

Finally, 
the numerical treatment of the problem
is simplified by  introducing a new (compact) radial
coordinate  $x$, such that $r=\frac{r_H}{1-x^2}$ and $0\leqslant x \leqslant 1$,
such that the horizon is located at $x=0$.
This results in the following boundary conditions at the horizon: 
\begin{eqnarray}
\label{eh}
&&\partial_x F_1=\partial_x F_2=\partial_x F_3=\partial_x F_0=0\ ,~ \nonumber \\
&&
S_1=S_2=\partial_x S_3 =0\ ,~V=0\ ,~\partial_x \phi=0\ .
\end{eqnarray}

A general large-$r$ expression of the solutions can
be constructed as a series in $1/r$ ($e.g.$ $F_0=1+c_t/r+\dots$, where $c_t$ is a constant).
The mass and the electric charge of the solutions
are read off from the far field asymptotics
\begin{equation}
\label{eqs1s}
-g_{tt}=F_0N=1-\frac{2M}{r}+\dots,~~V=\Phi+\frac{Q}{r}+\dots
\end{equation}
where $\Phi$ is the electrostatic potential.
Also, the solutions have an horizon area
\begin{equation}
\label{AHs1}
 A_H= r_H^2 \int_0^{2\pi} d\varphi \int_0^{\pi} d\theta  \sin\theta \sqrt{F_2  F_3} ~.
\end{equation}

We have studied in a systematic way the solutions with boundary data
corresponding to a
$\ell=1$, $m=0,1$ and 
$\ell=2$, $m=0,1,2$, for several values of the coupling constant $\alpha$.
The field equations are first discretized on a $(r, \theta,\varphi)$ grid with 
$N_r\times N_\theta \times N_\varphi$
points.
The grid spacing in the $r$ direction is non-uniform, while the values of the grid points in the angular
directions are uniform.
Typical grids have sizes around
$150 \times 24 \times 24$.
The resulting system is solved
iteratively until convergence is achieved. 
All numerical calculations for non-spherical configurations
are performed with a professional software~\cite{schoen}, based on the iterative Newton-Raphson
method. The code automatically provides an error estimate for each unknown function,
which is the maximum of the discretization error divided by the maximum of the function.
For most of the non-spherical solutions reported in this work,
 the typical numerical error is estimated to be around
$10^{-3}$.


\bibliography{letter_scalarised}

\begin{thebibliography}{43}
\expandafter\ifx\csname natexlab\endcsname\relax\def\natexlab#1{#1}\fi
\expandafter\ifx\csname bibnamefont\endcsname\relax
  \def\bibnamefont#1{#1}\fi
\expandafter\ifx\csname bibfnamefont\endcsname\relax
  \def\bibfnamefont#1{#1}\fi
\expandafter\ifx\csname citenamefont\endcsname\relax
  \def\citenamefont#1{#1}\fi
\expandafter\ifx\csname url\endcsname\relax
  \def\url#1{\texttt{#1}}\fi
\expandafter\ifx\csname urlprefix\endcsname\relax\def\urlprefix{URL }\fi
\providecommand{\bibinfo}[2]{#2}
\providecommand{\eprint}[2][]{\url{#2}}

\bibitem[{\citenamefont{Abbott et~al.}(2016{\natexlab{a}})}]{Abbott:2016blz}
\bibinfo{author}{\bibfnamefont{B.~P.} \bibnamefont{Abbott}}
  \bibnamefont{et~al.} (\bibinfo{collaboration}{Virgo, LIGO Scientific}),
  \bibinfo{journal}{Phys. Rev. Lett.} \textbf{\bibinfo{volume}{116}},
  \bibinfo{pages}{061102} (\bibinfo{year}{2016}{\natexlab{a}}),
  \eprint{1602.03837}.

\bibitem[{\citenamefont{Abbott et~al.}(2016{\natexlab{b}})}]{Abbott:2016nmj}
\bibinfo{author}{\bibfnamefont{B.~P.} \bibnamefont{Abbott}}
  \bibnamefont{et~al.} (\bibinfo{collaboration}{Virgo, LIGO Scientific}),
  \bibinfo{journal}{Phys. Rev. Lett.} \textbf{\bibinfo{volume}{116}},
  \bibinfo{pages}{241103} (\bibinfo{year}{2016}{\natexlab{b}}),
  \eprint{1606.04855}.

\bibitem[{\citenamefont{Abbott et~al.}(2017{\natexlab{a}})}]{Abbott:2017vtc}
\bibinfo{author}{\bibfnamefont{B.~P.} \bibnamefont{Abbott}}
  \bibnamefont{et~al.} (\bibinfo{collaboration}{VIRGO, LIGO Scientific}),
  \bibinfo{journal}{Phys. Rev. Lett.} \textbf{\bibinfo{volume}{118}},
  \bibinfo{pages}{221101} (\bibinfo{year}{2017}{\natexlab{a}}),
  \eprint{1706.01812}.

\bibitem[{\citenamefont{Abbott et~al.}(2017{\natexlab{b}})}]{Abbott:2017oio}
\bibinfo{author}{\bibfnamefont{B.~P.} \bibnamefont{Abbott}}
  \bibnamefont{et~al.} (\bibinfo{collaboration}{Virgo, LIGO Scientific}),
  \bibinfo{journal}{Phys. Rev. Lett.} \textbf{\bibinfo{volume}{119}},
  \bibinfo{pages}{141101} (\bibinfo{year}{2017}{\natexlab{b}}),
  \eprint{1709.09660}.

\bibitem[{\citenamefont{Abbott
  et~al.}(2017{\natexlab{c}})}]{TheLIGOScientific:2017qsa}
\bibinfo{author}{\bibfnamefont{B.}~\bibnamefont{Abbott}} \bibnamefont{et~al.}
  (\bibinfo{collaboration}{Virgo, LIGO Scientific}), \bibinfo{journal}{Phys.
  Rev. Lett.} \textbf{\bibinfo{volume}{119}}, \bibinfo{pages}{161101}
  (\bibinfo{year}{2017}{\natexlab{c}}), \eprint{1710.05832}.

\bibitem[{\citenamefont{Abbott et~al.}(2017{\natexlab{d}})}]{Abbott:2017gyy}
\bibinfo{author}{\bibfnamefont{B.~P.} \bibnamefont{Abbott}}
  \bibnamefont{et~al.} (\bibinfo{collaboration}{Virgo, LIGO Scientific}),
  \bibinfo{journal}{Astrophys. J.} \textbf{\bibinfo{volume}{851}},
  \bibinfo{pages}{L35} (\bibinfo{year}{2017}{\natexlab{d}}),
  \eprint{1711.05578}.

\bibitem[{\citenamefont{Kerr}(1963)}]{Kerr:1963ud}
\bibinfo{author}{\bibfnamefont{R.~P.} \bibnamefont{Kerr}},
  \bibinfo{journal}{Phys.Rev.Lett.} \textbf{\bibinfo{volume}{11}},
  \bibinfo{pages}{237} (\bibinfo{year}{1963}).

\bibitem[{\citenamefont{Carter}(1971)}]{Carter:1971zc}
\bibinfo{author}{\bibfnamefont{B.}~\bibnamefont{Carter}},
  \bibinfo{journal}{Phys. Rev. Lett.} \textbf{\bibinfo{volume}{26}},
  \bibinfo{pages}{331} (\bibinfo{year}{1971}).

\bibitem[{\citenamefont{Robinson}(1975)}]{Robinson:1975bv}
\bibinfo{author}{\bibfnamefont{D.~C.} \bibnamefont{Robinson}},
  \bibinfo{journal}{Phys. Rev. Lett.} \textbf{\bibinfo{volume}{34}},
  \bibinfo{pages}{905} (\bibinfo{year}{1975}).

\bibitem[{\citenamefont{Chrusciel et~al.}(2012)\citenamefont{Chrusciel, Costa,
  and Heusler}}]{Chrusciel:2012jk}
\bibinfo{author}{\bibfnamefont{P.~T.} \bibnamefont{Chrusciel}},
  \bibinfo{author}{\bibfnamefont{J.~L.} \bibnamefont{Costa}}, \bibnamefont{and}
  \bibinfo{author}{\bibfnamefont{M.}~\bibnamefont{Heusler}},
  \bibinfo{journal}{Living Rev.Rel.} \textbf{\bibinfo{volume}{15}},
  \bibinfo{pages}{7} (\bibinfo{year}{2012}), \eprint{1205.6112}.

\bibitem[{\citenamefont{Doneva and Yazadjiev}(2018)}]{Doneva:2017bvd}
\bibinfo{author}{\bibfnamefont{D.~D.} \bibnamefont{Doneva}} \bibnamefont{and}
  \bibinfo{author}{\bibfnamefont{S.~S.} \bibnamefont{Yazadjiev}},
  \bibinfo{journal}{Phys. Rev. Lett.} \textbf{\bibinfo{volume}{120}},
  \bibinfo{pages}{131103} (\bibinfo{year}{2018}), \eprint{1711.01187}.

\bibitem[{\citenamefont{Silva et~al.}(2018)\citenamefont{Silva, Sakstein,
  Gualtieri, Sotiriou, and Berti}}]{Silva:2017uqg}
\bibinfo{author}{\bibfnamefont{H.~O.} \bibnamefont{Silva}},
  \bibinfo{author}{\bibfnamefont{J.}~\bibnamefont{Sakstein}},
  \bibinfo{author}{\bibfnamefont{L.}~\bibnamefont{Gualtieri}},
  \bibinfo{author}{\bibfnamefont{T.~P.} \bibnamefont{Sotiriou}},
  \bibnamefont{and} \bibinfo{author}{\bibfnamefont{E.}~\bibnamefont{Berti}},
  \bibinfo{journal}{Phys. Rev. Lett.} \textbf{\bibinfo{volume}{120}},
  \bibinfo{pages}{131104} (\bibinfo{year}{2018}), \eprint{1711.02080}.

\bibitem[{\citenamefont{Antoniou et~al.}(2018)\citenamefont{Antoniou,
  Bakopoulos, and Kanti}}]{Antoniou:2017acq}
\bibinfo{author}{\bibfnamefont{G.}~\bibnamefont{Antoniou}},
  \bibinfo{author}{\bibfnamefont{A.}~\bibnamefont{Bakopoulos}},
  \bibnamefont{and} \bibinfo{author}{\bibfnamefont{P.}~\bibnamefont{Kanti}},
  \bibinfo{journal}{Phys. Rev. Lett.} \textbf{\bibinfo{volume}{120}},
  \bibinfo{pages}{131102} (\bibinfo{year}{2018}), \eprint{1711.03390}.

\bibitem[{\citenamefont{Blázquez-Salcedo
  et~al.}(2018)\citenamefont{Blázquez-Salcedo, Doneva, Kunz, and
  Yazadjiev}}]{Blazquez-Salcedo:2018jnn}
\bibinfo{author}{\bibfnamefont{J.~L.} \bibnamefont{Blázquez-Salcedo}},
  \bibinfo{author}{\bibfnamefont{D.~D.} \bibnamefont{Doneva}},
  \bibinfo{author}{\bibfnamefont{J.}~\bibnamefont{Kunz}}, \bibnamefont{and}
  \bibinfo{author}{\bibfnamefont{S.~S.} \bibnamefont{Yazadjiev}}
  (\bibinfo{year}{2018}), \eprint{1805.05755}.

\bibitem[{\citenamefont{Damour and Esposito-Farese}(1993)}]{Damour:1993hw}
\bibinfo{author}{\bibfnamefont{T.}~\bibnamefont{Damour}} \bibnamefont{and}
  \bibinfo{author}{\bibfnamefont{G.}~\bibnamefont{Esposito-Farese}},
  \bibinfo{journal}{Phys. Rev. Lett.} \textbf{\bibinfo{volume}{70}},
  \bibinfo{pages}{2220} (\bibinfo{year}{1993}).

\bibitem[{\citenamefont{Cardoso
  et~al.}(2013{\natexlab{a}})\citenamefont{Cardoso, Carucci, Pani, and
  Sotiriou}}]{Cardoso:2013fwa}
\bibinfo{author}{\bibfnamefont{V.}~\bibnamefont{Cardoso}},
  \bibinfo{author}{\bibfnamefont{I.~P.} \bibnamefont{Carucci}},
  \bibinfo{author}{\bibfnamefont{P.}~\bibnamefont{Pani}}, \bibnamefont{and}
  \bibinfo{author}{\bibfnamefont{T.~P.} \bibnamefont{Sotiriou}},
  \bibinfo{journal}{Phys. Rev. Lett.} \textbf{\bibinfo{volume}{111}},
  \bibinfo{pages}{111101} (\bibinfo{year}{2013}{\natexlab{a}}),
  \eprint{1308.6587}.

\bibitem[{\citenamefont{Hirschmann et~al.}(2018)\citenamefont{Hirschmann,
  Lehner, Liebling, and Palenzuela}}]{Hirschmann:2017psw}
\bibinfo{author}{\bibfnamefont{E.~W.} \bibnamefont{Hirschmann}},
  \bibinfo{author}{\bibfnamefont{L.}~\bibnamefont{Lehner}},
  \bibinfo{author}{\bibfnamefont{S.~L.} \bibnamefont{Liebling}},
  \bibnamefont{and}
  \bibinfo{author}{\bibfnamefont{C.}~\bibnamefont{Palenzuela}},
  \bibinfo{journal}{Phys. Rev.} \textbf{\bibinfo{volume}{D97}},
  \bibinfo{pages}{064032} (\bibinfo{year}{2018}), \eprint{1706.09875}.

\bibitem[{\citenamefont{Israel}(1968)}]{Israel:1967za}
\bibinfo{author}{\bibfnamefont{W.}~\bibnamefont{Israel}},
  \bibinfo{journal}{Commun. Math. Phys.} \textbf{\bibinfo{volume}{8}},
  \bibinfo{pages}{245} (\bibinfo{year}{1968}).

\bibitem[{\citenamefont{Bekenstein}(1995)}]{Bekenstein:1995un}
\bibinfo{author}{\bibfnamefont{J.~D.} \bibnamefont{Bekenstein}},
  \bibinfo{journal}{Phys. Rev.} \textbf{\bibinfo{volume}{D51}},
  \bibinfo{pages}{R6608} (\bibinfo{year}{1995}).

\bibitem[{\citenamefont{Herdeiro and Radu}(2015)}]{Herdeiro:2015waa}
\bibinfo{author}{\bibfnamefont{C.~A.~R.} \bibnamefont{Herdeiro}}
  \bibnamefont{and} \bibinfo{author}{\bibfnamefont{E.}~\bibnamefont{Radu}},
  \bibinfo{journal}{Int. J. Mod. Phys.} \textbf{\bibinfo{volume}{D24}},
  \bibinfo{pages}{1542014} (\bibinfo{year}{2015}), \eprint{1504.08209}.

\bibitem[{\citenamefont{Cardoso
  et~al.}(2013{\natexlab{b}})\citenamefont{Cardoso, Carucci, Pani, and
  Sotiriou}}]{Cardoso:2013opa}
\bibinfo{author}{\bibfnamefont{V.}~\bibnamefont{Cardoso}},
  \bibinfo{author}{\bibfnamefont{I.~P.} \bibnamefont{Carucci}},
  \bibinfo{author}{\bibfnamefont{P.}~\bibnamefont{Pani}}, \bibnamefont{and}
  \bibinfo{author}{\bibfnamefont{T.~P.} \bibnamefont{Sotiriou}},
  \bibinfo{journal}{Phys. Rev.} \textbf{\bibinfo{volume}{D88}},
  \bibinfo{pages}{044056} (\bibinfo{year}{2013}{\natexlab{b}}),
  \eprint{1305.6936}.

\bibitem[{\citenamefont{Garfinkle et~al.}(1991)\citenamefont{Garfinkle,
  Horowitz, and Strominger}}]{Garfinkle:1990qj}
\bibinfo{author}{\bibfnamefont{D.}~\bibnamefont{Garfinkle}},
  \bibinfo{author}{\bibfnamefont{G.~T.} \bibnamefont{Horowitz}},
  \bibnamefont{and}
  \bibinfo{author}{\bibfnamefont{A.}~\bibnamefont{Strominger}},
  \bibinfo{journal}{Phys. Rev.} \textbf{\bibinfo{volume}{D43}},
  \bibinfo{pages}{3140} (\bibinfo{year}{1991}), \bibinfo{note}{[Erratum: Phys.
  Rev.D45,3888(1992)]}.

\bibitem[{\citenamefont{Bekenstein}(1972)}]{Bekenstein:1972ny}
\bibinfo{author}{\bibfnamefont{J.~D.} \bibnamefont{Bekenstein}},
  \bibinfo{journal}{Phys. Rev. Lett.} \textbf{\bibinfo{volume}{28}},
  \bibinfo{pages}{452} (\bibinfo{year}{1972}).

\bibitem[{\citenamefont{Sanchis-Gual
  et~al.}(2016{\natexlab{a}})\citenamefont{Sanchis-Gual, Degollado, Montero,
  Font, and Herdeiro}}]{Sanchis-Gual:2015lje}
\bibinfo{author}{\bibfnamefont{N.}~\bibnamefont{Sanchis-Gual}},
  \bibinfo{author}{\bibfnamefont{J.~C.} \bibnamefont{Degollado}},
  \bibinfo{author}{\bibfnamefont{P.~J.} \bibnamefont{Montero}},
  \bibinfo{author}{\bibfnamefont{J.~A.} \bibnamefont{Font}}, \bibnamefont{and}
  \bibinfo{author}{\bibfnamefont{C.}~\bibnamefont{Herdeiro}},
  \bibinfo{journal}{Phys. Rev. Lett.} \textbf{\bibinfo{volume}{116}},
  \bibinfo{pages}{141101} (\bibinfo{year}{2016}{\natexlab{a}}),
  \eprint{1512.05358}.

\bibitem[{\citenamefont{Alcubierre}(2008)}]{Alcubierre08a}
\bibinfo{author}{\bibfnamefont{M.}~\bibnamefont{Alcubierre}},
  \emph{\bibinfo{title}{Introduction to $3+1$ Numerical Relativity}}
  (\bibinfo{publisher}{Oxford Univ. Press}, \bibinfo{address}{New York},
  \bibinfo{year}{2008}), ISBN \bibinfo{isbn}{978-0-19-920567-7}.

\bibitem[{\citenamefont{Headrick et~al.}(2010)\citenamefont{Headrick, Kitchen,
  and Wiseman}}]{Headrick:2009pv}
\bibinfo{author}{\bibfnamefont{M.}~\bibnamefont{Headrick}},
  \bibinfo{author}{\bibfnamefont{S.}~\bibnamefont{Kitchen}}, \bibnamefont{and}
  \bibinfo{author}{\bibfnamefont{T.}~\bibnamefont{Wiseman}},
  \bibinfo{journal}{Class. Quant. Grav.} \textbf{\bibinfo{volume}{27}},
  \bibinfo{pages}{035002} (\bibinfo{year}{2010}), \eprint{0905.1822}.

\bibitem[{\citenamefont{Adam et~al.}(2012)\citenamefont{Adam, Kitchen, and
  Wiseman}}]{Adam:2011dn}
\bibinfo{author}{\bibfnamefont{A.}~\bibnamefont{Adam}},
  \bibinfo{author}{\bibfnamefont{S.}~\bibnamefont{Kitchen}}, \bibnamefont{and}
  \bibinfo{author}{\bibfnamefont{T.}~\bibnamefont{Wiseman}},
  \bibinfo{journal}{Class. Quant. Grav.} \textbf{\bibinfo{volume}{29}},
  \bibinfo{pages}{165002} (\bibinfo{year}{2012}), \eprint{1105.6347}.

\bibitem[{\citenamefont{Dias et~al.}(2016)\citenamefont{Dias, Santos, and
  Way}}]{Dias:2015nua}
\bibinfo{author}{\bibfnamefont{O.~J.~C.} \bibnamefont{Dias}},
  \bibinfo{author}{\bibfnamefont{J.~E.} \bibnamefont{Santos}},
  \bibnamefont{and} \bibinfo{author}{\bibfnamefont{B.}~\bibnamefont{Way}},
  \bibinfo{journal}{Class. Quant. Grav.} \textbf{\bibinfo{volume}{33}},
  \bibinfo{pages}{133001} (\bibinfo{year}{2016}), \eprint{1510.02804}.

\bibitem[{\citenamefont{Ridgway and Weinberg}(1995)}]{Ridgway:1995ke}
\bibinfo{author}{\bibfnamefont{S.~A.} \bibnamefont{Ridgway}} \bibnamefont{and}
  \bibinfo{author}{\bibfnamefont{E.~J.} \bibnamefont{Weinberg}},
  \bibinfo{journal}{Phys. Rev.} \textbf{\bibinfo{volume}{D52}},
  \bibinfo{pages}{3440} (\bibinfo{year}{1995}), \eprint{gr-qc/9503035}.

\bibitem[{\citenamefont{Ioannidou et~al.}(2006)\citenamefont{Ioannidou,
  Kleihaus, and Kunz}}]{Ioannidou:2006mg}
\bibinfo{author}{\bibfnamefont{T.}~\bibnamefont{Ioannidou}},
  \bibinfo{author}{\bibfnamefont{B.}~\bibnamefont{Kleihaus}}, \bibnamefont{and}
  \bibinfo{author}{\bibfnamefont{J.}~\bibnamefont{Kunz}},
  \bibinfo{journal}{Phys. Lett.} \textbf{\bibinfo{volume}{B635}},
  \bibinfo{pages}{161} (\bibinfo{year}{2006}), \eprint{gr-qc/0601103}.

\bibitem[{\citenamefont{Herdeiro and Radu}(2016)}]{Herdeiro:2016plq}
\bibinfo{author}{\bibfnamefont{C.~A.} \bibnamefont{Herdeiro}} \bibnamefont{and}
  \bibinfo{author}{\bibfnamefont{E.}~\bibnamefont{Radu}},
  \bibinfo{journal}{Phys. Rev. Lett.} \textbf{\bibinfo{volume}{117}},
  \bibinfo{pages}{221102} (\bibinfo{year}{2016}), \eprint{1606.02302}.

\bibitem[{\citenamefont{Rayleigh}(1879)}]{doi:10.1112/plms/s1-11.1.57}
\bibinfo{author}{\bibfnamefont{L.}~\bibnamefont{Rayleigh}},
  \bibinfo{journal}{Proc. London Math. Soc.} \textbf{\bibinfo{volume}{10}},
  \bibinfo{pages}{4} (\bibinfo{year}{1879}).

\bibitem[{\citenamefont{Rayleigh}(1882)}]{doi:10.1080/14786448208628425}
\bibinfo{author}{\bibfnamefont{L.}~\bibnamefont{Rayleigh}},
  \bibinfo{journal}{The London, Edinburgh, and Dublin Philosophical Magazine
  and Journal of Science} \textbf{\bibinfo{volume}{14}}, \bibinfo{pages}{184}
  (\bibinfo{year}{1882}).

\bibitem[{\citenamefont{Basaran and Scriven}(1989)}]{doi:10.1063/1.857551}
\bibinfo{author}{\bibfnamefont{O.~A.} \bibnamefont{Basaran}} \bibnamefont{and}
  \bibinfo{author}{\bibfnamefont{L.~E.} \bibnamefont{Scriven}},
  \bibinfo{journal}{Physics of Fluids A: Fluid Dynamics}
  \textbf{\bibinfo{volume}{1}}, \bibinfo{pages}{795} (\bibinfo{year}{1989}).

\bibitem[{\citenamefont{Duft et~al.}(2003)\citenamefont{Duft, Achtzehn,
  M{\"u}ller, Huber, and Leisner}}]{nature}
\bibinfo{author}{\bibfnamefont{D.}~\bibnamefont{Duft}},
  \bibinfo{author}{\bibfnamefont{T.}~\bibnamefont{Achtzehn}},
  \bibinfo{author}{\bibfnamefont{R.}~\bibnamefont{M{\"u}ller}},
  \bibinfo{author}{\bibfnamefont{B.~A.} \bibnamefont{Huber}}, \bibnamefont{and}
  \bibinfo{author}{\bibfnamefont{T.}~\bibnamefont{Leisner}},
  \bibinfo{journal}{Nature} \textbf{\bibinfo{volume}{421}}, \bibinfo{pages}{128
  EP } (\bibinfo{year}{2003}).

\bibitem[{\citenamefont{Burton and Taborek}(2011)}]{PhysRevLett.106.144501}
\bibinfo{author}{\bibfnamefont{J.~C.} \bibnamefont{Burton}} \bibnamefont{and}
  \bibinfo{author}{\bibfnamefont{P.}~\bibnamefont{Taborek}},
  \bibinfo{journal}{Phys. Rev. Lett.} \textbf{\bibinfo{volume}{106}},
  \bibinfo{pages}{144501} (\bibinfo{year}{2011}).

\bibitem[{\citenamefont{Misner and Sharp}(1964)}]{Misner:1964je}
\bibinfo{author}{\bibfnamefont{C.~W.} \bibnamefont{Misner}} \bibnamefont{and}
  \bibinfo{author}{\bibfnamefont{D.~H.} \bibnamefont{Sharp}},
  \bibinfo{journal}{Phys. Rev.} \textbf{\bibinfo{volume}{136}},
  \bibinfo{pages}{B571} (\bibinfo{year}{1964}).

\bibitem[{\citenamefont{Sanchis-Gual
  et~al.}(2016{\natexlab{b}})\citenamefont{Sanchis-Gual, Degollado, Herdeiro,
  Font, and Montero}}]{Sanchis-Gual:2016tcm}
\bibinfo{author}{\bibfnamefont{N.}~\bibnamefont{Sanchis-Gual}},
  \bibinfo{author}{\bibfnamefont{J.~C.} \bibnamefont{Degollado}},
  \bibinfo{author}{\bibfnamefont{C.}~\bibnamefont{Herdeiro}},
  \bibinfo{author}{\bibfnamefont{J.~A.} \bibnamefont{Font}}, \bibnamefont{and}
  \bibinfo{author}{\bibfnamefont{P.~J.} \bibnamefont{Montero}},
  \bibinfo{journal}{Phys. Rev.} \textbf{\bibinfo{volume}{D94}},
  \bibinfo{pages}{044061} (\bibinfo{year}{2016}{\natexlab{b}}),
  \eprint{1607.06304}.

\bibitem[{\citenamefont{Torres and Alcubierre}(2014)}]{Torres:2014fga}
\bibinfo{author}{\bibfnamefont{J.~M.} \bibnamefont{Torres}} \bibnamefont{and}
  \bibinfo{author}{\bibfnamefont{M.}~\bibnamefont{Alcubierre}},
  \bibinfo{journal}{Gen. Rel. Grav.} \textbf{\bibinfo{volume}{46}},
  \bibinfo{pages}{1773} (\bibinfo{year}{2014}), \eprint{1407.7885}.

\bibitem[{\citenamefont{Schonauer and Weiss}(1989)}]{schoen}
\bibinfo{author}{\bibfnamefont{W.}~\bibnamefont{Schonauer}} \bibnamefont{and}
  \bibinfo{author}{\bibfnamefont{R.}~\bibnamefont{Weiss}}, \bibinfo{journal}{J.
  Comput. Appl. Math.} \textbf{\bibinfo{volume}{27}}, \bibinfo{pages}{279}
  (\bibinfo{year}{1989}).

\bibitem[{\citenamefont{Gubser}(2005)}]{Gubser:2005ih}
\bibinfo{author}{\bibfnamefont{S.~S.} \bibnamefont{Gubser}},
  \bibinfo{journal}{Class. Quant. Grav.} \textbf{\bibinfo{volume}{22}},
  \bibinfo{pages}{5121} (\bibinfo{year}{2005}), \eprint{hep-th/0505189}.

\bibitem[{\citenamefont{Stefanov et~al.}(2008)\citenamefont{Stefanov,
  Yazadjiev, and Todorov}}]{Stefanov:2007eq}
\bibinfo{author}{\bibfnamefont{I.~Z.} \bibnamefont{Stefanov}},
  \bibinfo{author}{\bibfnamefont{S.~S.} \bibnamefont{Yazadjiev}},
  \bibnamefont{and} \bibinfo{author}{\bibfnamefont{M.~D.}
  \bibnamefont{Todorov}}, \bibinfo{journal}{Mod. Phys. Lett.}
  \textbf{\bibinfo{volume}{A23}}, \bibinfo{pages}{2915} (\bibinfo{year}{2008}),
  \eprint{0708.4141}.

\bibitem[{\citenamefont{Doneva et~al.}(2010)\citenamefont{Doneva, Yazadjiev,
  Kokkotas, and Stefanov}}]{Doneva:2010ke}
\bibinfo{author}{\bibfnamefont{D.~D.} \bibnamefont{Doneva}},
  \bibinfo{author}{\bibfnamefont{S.~S.} \bibnamefont{Yazadjiev}},
  \bibinfo{author}{\bibfnamefont{K.~D.} \bibnamefont{Kokkotas}},
  \bibnamefont{and} \bibinfo{author}{\bibfnamefont{I.~Z.}
  \bibnamefont{Stefanov}}, \bibinfo{journal}{Phys. Rev.}
  \textbf{\bibinfo{volume}{D82}}, \bibinfo{pages}{064030}
  (\bibinfo{year}{2010}), \eprint{1007.1767}.

\end{thebibliography}

 
\end{document}